\documentstyle [aps,prb,twocolumn,epsf]{revtex}

\newcommand{\br}{{\bf r}}
\newcommand{\bq}{{\bf q}}
\newcommand{\be}{\begin{equation}}
\newcommand{\ee}{\end{equation}}
\newcommand{\bea}{\begin{eqnarray}}
\newcommand{\eea}{\end{eqnarray}}
\newcommand{\ba}{\begin{array}}
\newcommand{\ea}{\end{array}}

\begin{document}
\draft

\title{
Effect of Landau Level Mixing for Electrons in Random Magnetic Field
}

\author {Ming-Che Chang,${}^1$ Min-Fong Yang,${}^2$ and Tzay-Ming Hong$^1$}
\address{${}^1$
Department of Physics, National Tsing Hua University,
Hsinchu, Taiwan\\
${}^2$ Department of General  Programs,
Chang Gung College of Medicine and Technology,
Kweishan, Taoyuan, Taiwan}

\address{\mbox{ }}
\address{\parbox{14cm}{\rm \mbox{ }\mbox{ }
An effective Hamiltonian approach is used to study the effect of
Landau-level mixing on the energy spectrum of electrons
in a smooth but random magnetic field $B(\br)$ with a {\it finite}
uniform component $B_0$. It is found that, as opposed to electrostatic
disorder, the energy levels of localized electrons shift {\it
upward} with a rate of order ${\cal O}(B_0^{-1})$ when $B_0$ is
decreased, while
the extended states remain static at the same order. Therefore, there is
no indication that the
extended states will float out of the Fermi energy and induce a
metal-insulator transition as the magnetic disorder is increased.
We also find that the Zeeman term may have significant effect on the
spectral shift of low-lying Landau levels.
}}
\address{\mbox{ }}
\address{\parbox{14cm}{\rm PACS numbers: 73.40.Hm, 71.30.+h}}
\date{\today}

\maketitle
Recently, there is intensive interest on the
problem of two-dimensional electron gas in a static random
magnetic field (RMF). First, this problem is
related to the localization problem for the ``composite fermions''
in the fractional quantum Hall effect.\cite{Jain}
In the mean-field treatment, the composite fermions
move in a weak effective magnetic field that contains a
random component induced by the inhomogeneous electron density.
Second, the study of RMF may be applied to high-T$_c$
superconductivity systems, where the RMF is considered as a limiting
case of the gauge-field.\cite{HTc} Third, recent experiments that
measured transport properties of electrons in a static RMF
also add considerable interest to this subject.\cite{exps}

Most theoretical studies in the literature focused on free electrons
in a RMF with
{\it zero mean value}. A central issue is whether all the electrons are
localized in such an environment. The
results are rather controversial. Analytically, due to the zero
average of the magnetic field, the field-theoretical description
corresponds to a non-linear
sigma model of the  unitary class without a topological term, which predicts
that all states are localized,\cite{AMW} according to the conventional
scaling theory.\cite{AALR}
However, Zhang and Arovas \cite{ZA} have
suggested that a long-range logarithmic interaction  between the
topological densities (due to the local fluctuations of the Hall conductance)
may lead to delocalization.
The numerical works on a finite lattice only add more conflicting
results.  Some authors
claim that there may exist the mobility edge separating
the localized states from the extended states; \cite{KZ,KWAZ}
however, other authors, while observing a strong
enhancement of the localization length, find no true transition.
\cite{SN,LCK}
The controversy in the numerical works arises from the interpretation of
their data. Because the localization length
increases rapidly as a function of energy when the band
center is approached, it is hard to distinguish whether the states are
really extended or weakly localized with
the localization length much longer than the sample size.

On the other hand, our understanding of electron localization in
a random electric potential is more complete.\cite{HuckR} It is known that
there are extended states at the centers of Landau bands in a strong uniform
magnetic field. In order to be consistent with the conventional scaling
theory for the zero-field case,\cite{AALR} it is argued that the
extended states will float up in energy if the magnetic
field strength is reduced (or equivalently, if the strength of disorder
is increased),\cite{float}  and therefore all the states below the Fermi
energy are eventually localized. Although this levitation scenario is
appealing, its microscopic foundation is not clear. Recently, by using a
simple perturbative approach, Haldane and Yang \cite{HY} show that the
levitation of the extended states can be explained as a result of
Landau-level mixing, thus support the levitation scenario.

Motivated by the work of Haldane and Yang,\cite{HY}
we study the spectral shift of the two-dimensional
electrons in a RMF $B({\bf r})=B_0 + b({\br})$ when its spatial average $B_0$
is reduced. As pointed out by Kalmeyer {\it et al.}\cite{KWAZ} (see also
Ref.~\onlinecite{Huck96}), when $B_0 \neq 0$ and $b({\br}) \ll B_0$,
the random fluctuation behaves like a random scalar
potential. In this case, one recovers the well-studied problem of
electrons in a random potential and a uniform magnetic field,
thus it is expected that there are extended states at the centers of
Landau bands.
If the correlation length of the disorder is much longer than
the magnetic length $\ell=\sqrt{\hbar/eB_0}$,
the motion of electrons can be decomposed
into a fast cyclotron motion and a slow guiding-center motion.\cite{LCK}  The
guiding centers move along the contours of $b({\bf r})$ with the local
drift velocity ${\bf V}_{\rm d}=(e\xi^2/2m)\nabla b \times {\bf \hat
z}$, where $\xi$ is the cyclotron radius and $m$ is the
electron mass.\cite{Jackson} (See Fig.~1.) Around hills or valleys of $b({\bf
r})$, the contours are closed and the corresponding states are localized. The
extended states occur only at the percolation contour
whose energy is determined by the saddle points of $b({\br})$, similar
to the semiclassical theory for electrostatic disorder.
Due to this similarity, Lee {\it et al.}\cite{LCK}
propose that the extended states will levitate in energy
with decreasing $B_0$, and hence all states below the Fermi energy should
be localized when $B_0 = 0$. However, by using the same perturbative
approach used in Ref.~\onlinecite{HY}, we find that
the leading term of the effective Hamiltonian will cause the
{\it localized states}, rather than the extended states, to float up in
energy as $B_0$ decreases. Thus, the levitation scenario of extended
states in the RMF case has {\it no} firm support, and, therefore, gives {\it
no} implication to electron localization in a RMF.
Furthermore, we show that the Zeeman term may have significant
effect on the spectral shift of low-lying Landau levels.

For the two-dimensional electron gas in a RMF
with a {\it nonzero} average, the Hamiltonian $H$ is composed of three parts,
\begin{eqnarray}
    H_0 &=& \frac{1}{2m}\left({\bf p}+e{\bf A}\right)^2,\\
    H_1 &=& \frac{e}{2m} \left[ ({\bf p}+e{\bf A}) \cdot {\bf a}
        +{\bf a} \cdot ({\bf p}+e{\bf A}) \right],\\
    H_2 &=& \frac{e^2}{2m} {\bf a}^2,
\label{H2}
\end{eqnarray}
where ${\bf A}$ and ${\bf a}$ are the vector potentials for $B_0{\bf \hat
z}$ and  $b({\bf r}) {\bf \hat z}$, respectively.
Using the Coulomb gauge for the fluctuating vector potential,
we can write
\begin{equation}
{\bf a}({\bf r})
= {1 \over {\cal A}} \sum_{{\bf q} \neq 0} i {\bf q} \times {\bf \hat z}
             { b({\bf q}) \over q^2}  e^{i{\bf q}\cdot {\bf r}},
\label{Four}
\end{equation}
where $b({\bf q})$ is the Fourier components of $b({\bf r})$, and
(quasi-)periodic boundary condition is imposed on the area ${\cal A}$ that
contains an integer number of magnetic flux quanta.
For a smooth and weak disorder, it is convenient to decompose the
position of an electron into a fast cyclotron motion
${\bf r-R}$ and a slow guiding-center motion
${\bf R} = \left( x - \Pi_y /eB_0 \right) {\bf \hat x} +
          \left( y + \Pi_x /eB_0 \right) {\bf \hat y}$,  where
${\bf \Pi} = {\bf p} + e{\bf A}$ is the canonical momentum
operator. It can be shown that the fast and the slow parts commute with
each other and decouple nicely.\cite{Mac}
The velocity of the guiding center at the $n$-th Landau level can be
obtained by the Heisenberg equation of motion.
To lowest order, the result is
\begin{eqnarray}
\label{gc}
\frac{d}{dt} \langle n | {\bf R} | n \rangle
&=& \frac{1}{i\hbar}\langle n |[{\bf R},H] | n \rangle \cr
&\simeq& \frac{e}{m}\left(n+\frac{1}{2}\right)\ell^2
       \langle n | \nabla b \times {\bf \hat z} | n \rangle.
\end{eqnarray}
This form coincides with the classical expression $(e\xi^2/2m) \nabla b
\times {\bf \hat z} $, since the cyclotron radius $\xi$ at the $n$-th Landau
level is given by $\sqrt{ \langle n | ({\br} -{\bf R})^2 | n \rangle }
=\sqrt{2n+1}\ell$.

In the presence of the random field $b({\bf r})$, the Landau level index
$n$ is no longer a good quantum number and different
levels couple with each other. If the magnetic field $B_0$ is
very strong, we need only consider the projected Hamiltonian
in the subspace of a given Landau level. However, in
general, (virtual) transitions between different Landau levels will
renormalize the potential seen by the
electrons in this Landau level. By the perturbative renormalization in terms of
powers of $\bf a$, the effective
Hamiltonian for electrons in the $n$-th Landau band can be written as
\begin{equation}
\langle n | H^{(n)}_{\rm eff}({\bf r}) | n \rangle
= \left( n + \frac{1}{2} \right) \hbar \Omega
+ \sum_{k\ge 1} \langle n | V_k^{(n)}({\bf r}) | n \rangle ,
\label{effH}
\end{equation}
where $\hbar \Omega = \hbar e B_0 / m$ is the cyclotron energy
and $|n\rangle$ is the eigenstate of $H_0$. The effective
potential propotional to $\bf a$ is
\begin{equation}
\langle n|V^{(n)}_1({\bf r})|n\rangle = \langle n | H_1 | n \rangle,
\end{equation}
and the correction that is quadratic in $\bf a$ is given by
\begin{equation}
\langle n|V^{(n)}_2({\bf r})|n \rangle
= \langle n | H_2 | n \rangle +
{\sum_{n'\neq n}} { \langle n | H_1 | n' \rangle \langle n' | H_1 | n \rangle
              \over
              \hbar \Omega (n-n')}.
\label{v2}
\end{equation}

To first order, a direct calculation yields \cite{KWAZ,Huck96,CN}
\begin{equation}
\label{eh}
V_1^{(n)}({\br})
= -\frac{e \hbar}{m \ell^2} {1\over {\cal A}}
\sum_{{\bf q} \neq 0}
      { b({\bf q}) \over q^2 } g^{(n)}({\bq})
      e^{i {\bf q} \cdot {\br} },
\end{equation}
where
\be
g^{(n)}({\bq})=
\frac{\frac{\partial}{\partial\lambda} U_{nn}({\bf q}\lambda)|_{\lambda = 1}}
     {U_{nn}({\bf q}) },
\ee
in which $U_{nn}(\bq)$ is the diagonal part of
$U_{nn'}({\bq})=\langle n | e^{i{\bf q}\cdot ({\bf r}-{\bf R})} | n' \rangle$.
For brevity, the projection by $|n\rangle$ and its adjoint will be
neglected from now on. Bear in mind that the equality holds only in the
projected subspace of the $n$-th Landau level.
For the slowly varying $b({\bf r})$ (compared to the magnetic length),
only the small ${\bf q}$ components in Eq.~(\ref{eh}) make
significant contribution, hence
one can expand $g^{(n)}({\bq})$ into a power series in $q\ell$.
Up to the order of $(q\ell)^4$ for
$g^{(n)}({\bq})$, $V_1^{(n)}({\br})$ can be written as \cite{note}
\begin{equation}
V_1^{(n)} ({\br})
\simeq \left( n+\frac{1}{2}\right )\frac{\hbar e}{m}b({\bf r})
-\frac{n(n+1)}{4}\frac{\hbar e}{m}\ell^2 \nabla^2 b({\bf r}).
\label{1st}
\end{equation}
Both terms in Eq.~(\ref{1st}) lead to broadening of the Landau levels:
The first term lifts the energy degeneracy for electrons drifting along
different contours of $b({\br})$; the second term gives a positive
(negative) contribution to energy for electrons drifting along hills
(valleys) of $b({\br})$, which have negative (positive) curvature of
$b({\br})$, thus broadens the level further.
However, neither gives a net shift to the overall profile of the density
of states.

By using the algebra of $\bf \Pi$, the
second term of $V^{(n)}_2({\bf r})$ can be expressed as
\begin{equation}
\frac{e^2}{m\ell^2} {1\over {\cal A}^2}
\sum_{{\bf q} \neq 0, {\bf q}' \neq 0}
      { b({\bf q}) \over q^2 } { b({\bf q}') \over {q'}^2 }
      f^{(n)}({\bf q},{\bf q'})
      e^{i ( {\bf q} + {\bf q}') \cdot {\br} },
\end{equation}
where
\begin{eqnarray}
f^{(n)}(\bq,\bq')&=&\frac{e^{-i {\bf \hat z}\cdot\bq\times\bq'\ell^2/2}}
{U_{nn}(\bq+\bq')}\cr
&\times& {\sum_{n' \neq n}}
\frac{\frac{\partial}{\partial\lambda} U_{nn'}({\bf q}\lambda)|_{\lambda = 1}
\frac{\partial}{\partial\lambda} U_{n'n}({\bf q}'\lambda)|_{\lambda =
1}} {n-n'}.
\end{eqnarray}
The expansion of $f^{(n)}(\bq,\bq')$ in powers of $q\ell$ is given by
\begin{eqnarray}
f^{(n)}({\bf q},{\bf q'})
&=&
\frac{{\bf q} \cdot {\bf q'}}{2}\ell^2
- \left( n + \frac{1}{2} \right)
 \frac{{\bf q} \cdot {\bf q'}}{2}(q^2 + {q'}^2) \ell^4\cr
&+&{\cal O}(q^6\ell^6).
\label{fn}
\end{eqnarray}
The first term in Eq.~(\ref{fn}) gives $-(e^2/2m){\bf a}^2$ to
$V^{(n)}_2({\bf r})$, which
is negative for all states and cancels the first term in Eq.~(\ref{v2}).
Thus, up to the order of $(q\ell)^4$ for
$f^{(n)}({\bf q},{\bf q'})$,
\begin{equation}
V^{(n)}_2({\bf r})
\simeq \frac{e^2}{m} \left( n + \frac{1}{2} \right) \ell^2
        {\bf a} \cdot \nabla b \times {\bf \hat z}.
\label{v2_2}
\end{equation}

Notice that the effective potential $V^{(n)}_2({\bf r})$ is not
manifestly gauge invariant.\cite{Kohn}
However, this lack of gauge invariance does not
appear in the energy expectation value of the electronic states.
Under the semiclassical approximation, for electrons circling a closed orbit
${\cal C}$ with a constant energy ${\cal E}$, the energy
expectation value altered by $V^{(n)}_2({\bf r})$ is proportional to the
following integral:
\begin{eqnarray}
\langle V^{(n)}_2 \rangle
&\propto& \int d^2{\br} \delta({\cal E}^{(n)}({\br})-{\cal E})
         \frac{e^2}{m}\left(n+\frac{1}{2}\right)\ell^2
         {\bf a} \cdot \nabla b \times {\bf \hat z} \cr
&=& \oint_{\cal C}
    \frac{d l}{|\nabla {\cal E}^{(n)}({\br})|}
    \frac{e^2}{m}\left(n+\frac{1}{2}\right)\ell^2
    {\bf a} \cdot \nabla b \times {\bf \hat z} \cr
&\simeq& \frac{1}{B_0} \oint_{\cal C} d{\bf l} \cdot {\bf a},
\end{eqnarray}
where we have used the fact that the local energy of the $n$-th Landau band
${\cal E}^{(n)}({\br}) \simeq
(n+1/2) \hbar \Omega + (n+1/2)(\hbar e/m) b({\br})$
(see Eqs.~(\ref{effH}) and (\ref{1st})),
and $d{\bf l}$ is in the direction of ${\bf V}_d$.
That is, $\langle V^{(n)}_2 \rangle$ is proportional to the magnetic flux of
$b({\br})$ enclosed by ${\cal C}$ and is {\it positive} for
{\it both} of the orbits circling the hill and the valley. (See Fig.~1.)
Hence $\langle V^{(n)}_2 \rangle$ is gauge invariant
(as it should be) and gives a {\it upward} shift in energy for the localized
states. For the extended states, the shift is determined by the saddle
points of $b({\br})$,\cite{LCK} where $\nabla b({\bf r})=0$.
Therefore, $V^{(n)}_2({\bf r})$ vanishes (thus
also gauge-invariant), and the energy of the extended states remains
static at this order.

\begin{figure}
\epsfxsize=3.3truein
\epsffile{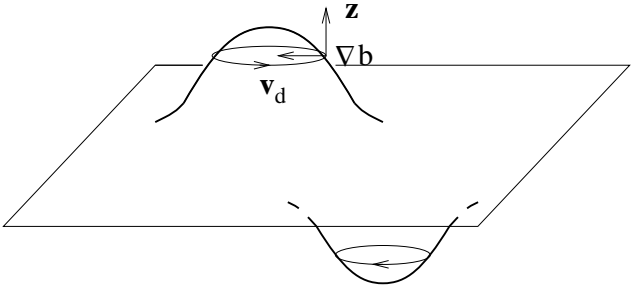}
\vspace{.8 cm}
\caption{Schematic diagram of the guiding-center orbits of
electrons circling around the hill and valley of the random magnetic
field. Note that the
sense of rotation is opposite for the two paths in the figure.}
\end{figure}

It is quite interesting to compare our result with that
of the electrostatic disorder case. \cite{HY} For electrostatic
disorder, it is found that the energies of the localized states shift
downward and that of
the extended states is static at order ${\cal O}(B_0^{-2})$.\cite{HY}
The downward movement is a manifestation of the generic
``level-repulsion" effect at the second order perturbation.
At the order of ${\cal O}(B_0^{-3})$, which is from the $(q\ell)^4$ term 
at the second order perturbation, the 
energy of the extended states shifts upward in stronger disorder and this 
behavior supports the levitation scenario\cite{float} to explain the 
metal-insulator transition.
However, the spectral shift in the RMF case is very different:
the energies of the localized states shift {\it upward}, and that of
the extended states remains static at the same order.
In a relative sense, the extended states move downward with respect to the
other states. Therefore, it seems unlikely that the extended states
will float out of the Fermi energy at strong disorder and induce a
metal-insulator transition. It might appear that the result
presented here contradicts the generic level-repulsion effect, which would
result in lowering of the levels (especially the lowest Landau level).
This is not so. The level-repulsion effect due to the level mixing should come
from the second term in Eq.~(\ref{v2}). As indicated in Eq.~(\ref{fn}),
the leading contribution $-(e^2/2m){\bf a}^2$ indeed
contributes to downward movement. However, this downward movement is
canceled by the diamagnetic term $(e^2/2m) {\bf a}^2$ that comes from
the first term in Eq.~(\ref{v2}). This cancellation is unique in the
magnetic disorder problem.

In the following, we would like to discuss briefly the influence of
the Zeeman term on the spectral shift. Besides contributing a
constant shift in energy, $\pm (g/4)\hbar\Omega$ ($g$ is the
electron $g$-factor),
as it does for the electrostatic disorder problems, the Zeeman term adds a
$b({\br})$-dependent part $H_z=-(g\hbar e/4m) \sigma_3 b({\br})$ to Eq.~(2),
where $\sigma_3$ is the Pauli matrix.
Consequently, the inclusion of the
Zeeman term leads to the following changes in the perturbative
calculation:\cite{note2}
for the first order calculation, we get an extra term
$-(g\hbar e/4m)\sigma_3b({\br})$ to $V_1^{(n)}({\br})$; while the additional
contribution to $V_2^{(n)}({\br})$ is given by
\be
{\sum_{n' \neq n}}
\left\{ {\langle n | H_1 | n' \rangle \langle n' | H_z | n  \rangle  \over
              \hbar \Omega (n-n')}+{\rm h.c.} \right\}
+{\sum_{n'\neq n}} { \langle n | H_z | n' \rangle \langle n' | H_z | n \rangle
              \over
              \hbar \Omega (n-n')}.
\ee
A straightforward calculation shows that the first term in the equation
above contributes
$-(ge^2/4m)\sigma_3 {\bf a} \cdot \nabla b \times {\bf \hat z}$ to
Eq.~(\ref{v2_2}); while the second term is of higher order in $q\ell$ and
can be neglected. Note that, apart from a multiplicative constant, this term
has the same form as the term in Eq.~(\ref{v2_2}). Consequently,
the conclusion that the extended states are not shifted, because
$\nabla b(\br)=0$ at the saddle points, remains valid.
Also note that the additional contribution is dependent on spin
but independent of $n$.
Therefore, the spectrum may shift differently between low-lying states
and higher levels: for Landau levels with
$(n+1/2) > g/4$, the localized states always move upward, but it may
become {\it downward} for spin-{\it up} electrons, if $(n+1/2) < g/4$.
In particular, for spin-up electrons at the lowest Landau level (LLL),
if $g=2$, then the $b(\br)$-dependent effective potentials in
Eqs.~(\ref{1st}) and (\ref{v2_2}) are canceled by these extra terms
due to the Zeeman term. If fact, it is not
difficult to prove that the cancellation is exact to all orders of $q\ell$
in $V_1^{(n)}({\br})$ and $V_2^{(n)}({\br})$.
This cancellation is consistent with the Aharonov-Casher theorem,\cite{AC}
which states that the LLL of spin-up electrons with $g=2$ will not be
broadened by magnetic disorder, no matter how strong the disorder is.

Finally, some comments are in order:
First, our result seems to be against the proposal that all states below the
Fermi energy are localized when $B_0=0$.\cite{AMW,SN,LCK} However,
since our perturbative approach is valid only for weak $b({\br})$
(compared to $B_0$), it is not sufficient to predict whether the extended
states will remain static when $B_0 \to 0$ and become the delocalized states
suggested in Refs.~\onlinecite{ZA,KZ,KWAZ}. Therefore,
to settle down the localization problem for the $B_0 = 0$ case, an alternative
approach that is applicable to the $B_0 \ll 1$ limit is urgently needed.
Second, the calculation presented here may be related to the $1/3 \to
1/2$ transition of the quantum Hall systems (i.e., the $1 \to 0$
transition of the composite fermions) by tuning the external field at a
given magnetic disorder --- if the ubiquitous electrostatic disorder in real
systems does not dominate the spectral shift. As mentioned above, depending
on the magnitude of the $g$ factor, the Zeeman term may lead to
different spectral shift between spin-up and spin-down electrons. It
would be interesting to observe this subtle behavior in future experiments.

\acknowledgments
The authors would like to thank C.~Y.~Mou for many valuable
discussions. This work is supported by the National Science Council of
Taiwan under contract No. NSC 86-2112-M-007-026 and 86-2112-M-002-028.


\begin{references}
\bibitem{note}
We believe that, in Eq.~(2.10) of Ref.~\onlinecite{KWAZ}, the authors miss
a factor 2 in
the coefficients of the second and the third terms of their gradient expansion.
\bibitem{note2} Notice that
the constant shift in energy $-\sigma_3(g/4)\hbar\Omega$ is added to
$H_0$ and
therefore gives a correction to the first term of Eq.~(\ref{effH}).
\end{references}
\end{document}